# Molding the flow of light at deep sub-wavelength scale

**Seunghoon Han, Yi Xiong, Dentcho Genov, Zhaowei Liu, Guy Bartal, and Xiang Zhang\***

NSF Nano-scale Science and Engineering Center (NSEC), University of California, Berkeley,

California, 94720, USA

\* e-mail: xiang@berkekely.edu

The diffractive nature of light[1-3] has limited optics and photonics to operate at scales much larger than the wavelength of light. The major challenge in scaling-down integrated photonics is how to mold the light flow below diffraction-limit in all three dimensions (Fig 1a). A high index solid immersion lens[4] can improve the spatial resolution by increasing the medium refractive index, but only to few times higher than in air. Photonic crystals[5-8] can guide light in three dimensions, however, the guided beam width is around a wavelength. Surface plasmons[9-11] has a potential to reach the sub-wavelength scales; nevertheless, it is confined in the two-dimensional interface between metals and dielectrics. Here, we present a new approach for molding the light flow at the deep sub-wavelength scale, using metamaterials with uniquely designed dispersion. We develop a design methodology for realizing *sub-wavelength ray optics*, and demonstrate λ/10 width light beams flow through three-dimensional space.

Ray optics is a powerful tool for designing and analyzing optical systems by tracing the ray trajectories in the medium according to a simple set of geometrical rules[1,2]. It is an ideal form of light propagation, where wave-optics phenomena such as diffraction of a narrow beam or wavefront distortion due to variations in the material response are assumed to be negligible. Ray optics is based on the "short wavelength approximation": the characteristics lengths associated with both diffraction and material variations are very large comparing to the wavelength. Consequently, the flow of light in a three-dimensional space can be realized by mapping the direction of light rays at each location, resulting in an orthogonal relationship between rays and geometrical wavefronts[1].

Notwithstanding its significant contribution for numerous applications, ray optics fails to describe light propagation at the scale of or below a wavelength. The elliptic dispersion curve in naturally occurring materials limits the range of propagating modes in momentum space, and results in a strong diffraction of light, as different constituents of a narrow beam propagate with different phase velocities. Moreover, to facilitate light propagation along a wavelength-scale curved trajectory, the host material should have rapidly varying optical response, resulting in a wavefront distortion. Consequently, the basic assumptions for ray optics are not valid in this setting. Molding the light flow in sub-wavelength scale by means of ray optics necessitates the elimination of diffraction of sub-wavelength light beams, and tailoring the material properties such that the *effective* optical response varies slowly along propagation.

The emerging field of *metamaterials*[12-14] has recently provided such means to design artificial materials with unusual optical properties that may not be found in their naturally occurring counterparts. The unit cells in these electromagnetic composites are typically much smaller than the wavelength, imposing material properties that can be ascribed as effective homogeneous medium[15, 16]. Metamaterials has shown extraordinary functionalities, such as negative refraction[3, 17], cloaking of arbitrary objects[18-19] and superlens[20-22], and may pave the way to control the diffraction and dispersion of light beams at will.

In this Letter, we introduce the concept of *sub-wavelength ray optics* in a new class of metamaterials having *locally varying* dispersion that is designed to be almost flat over a large range in momentum space ($k_x, k_y \gg 2\pi/\lambda$, $\lambda$ is the wavelength in vacuum).

The large range of propagating modes in momentum space allow the propagation of light beams at sub-wavelength scale, while the quasi-flat dispersion, ensures that these beams do not diffract in the effective medium, as all their constituents can propagate with the same phase velocity. The rapidly varying local material response can be engineered such that the effective response is slowly varying in a transformed space. As a result, ray-like light flow becomes feasible in the deep sub-wavelength scale similarly to ray optics in much larger scales[1, 2]. Using this design methodology, we realize a light beam of λ/10 width flow in 3D space. The efficiency of such design methodology is verified by a full wave simulation of propagation.

We start with a general physical picture in which the concepts of ray optics can be introduced to the sub-wavelength scale. Metamaterial with rapidly-varying optical properties can still comply with the ray optics approximations under a certain condition; the existence of an equivalent space, in which the light experiences an effective homogenous medium along its propagation, and flows along *linear* ray trajectories. The invariance of Maxwell equation under *conformal transformation* [23] ensures that the ray-like behavior will be sustained also along the original space, in which the light propagates along *curved* ray trajectories (Fig. 1b).

In what follows, we introduce the equivalent 2D space, obtained by a conformal transformation, and represented by the coordinate system $(u,v)$, which can be easily extended to a three-dimensional space by translation or rotation[23]. The light flows in the form of ray trajectories along the *v*-axis (Figure 1b). We define $\vec{s} = (\varepsilon, \mu)$ and $\vec{s}' = (\varepsilon', \mu')$ to be the diagonalized constituents of the permittivity and permeability along the $(u,v,w)$ axes in the Cartesian and curvilinear representations of $(u,v,w)$, respectively (see figs 1 and 2 in the Supplementary Information). The relations between $\vec{s}$ and $\vec{s}'$ are given by

$$s'_u = s_u \frac{1}{h_w}, \quad s'_v = s_v \frac{1}{h_w}$$
$$s'_w = s_w \frac{h_w}{h_u h_v} \tag{1}$$

where

$$h_u^2 = h_v^2 = \left(\frac{\partial x}{\partial u}\right)^2 + \left(\frac{\partial y}{\partial u}\right)^2, \quad h_w^2 = \left(\frac{\partial z}{\partial w}\right)^2 \tag{2}$$

are the Lame coefficients[18]. The degeneracy ($h_u = h_v$) originates from the Cauchy-Riemann relationships ($\partial x/\partial u = \partial y/\partial v$, $\partial x/\partial v = -\partial y/\partial u$)[23] and simplifies the design and analysis of the optical system.

A diffraction-free propagation throughout the medium requires dispersion relations ($k_v(k_u)$ where *v*-axis is the light flow direction) in the form of

$$k_v^2 = \varepsilon'_w \mu'_u \left( \frac{\omega^2}{c^2} - \frac{1}{\varepsilon'_w \mu'_v} k_u^2 \right), \qquad \varepsilon'_w \mu'_u = \varepsilon_w \mu_u / h_u^2 > 0 \text{ and}$$

$$(\varepsilon'_w \mu'_v = \varepsilon_w \mu_v / h_u^2 \to +\infty \text{ or } \mu'_u / \mu'_v = \mu_u / \mu_v \to -0) \qquad (3a)$$

for TE (electric field along $w$ axis) and

$$k_v^2 = \varepsilon'_u \mu'_w \left( \frac{\omega^2}{c^2} - \frac{1}{\varepsilon'_v \mu'_w} k_u^2 \right), \qquad \varepsilon'_u \mu'_w = \varepsilon_u \mu_w / h_u^2 > 0 \text{ and}$$

$$(\varepsilon'_v \mu'_w = \varepsilon_v \mu_w / h_u^2 \to +\infty \text{ or } \varepsilon'_u / \varepsilon'_v = \varepsilon_u / \varepsilon_v \to -0) \qquad (3b)$$

for TM (magnetic field along $w$ axis) polarizations.[15, 16]

Namely, the propagation constant $k_v$ should be real and non varying over a large range in the transverse momentum, in both the Cartesian and the curvilinear representations of $(u, v, w)$ (see fig 1c). This condition can be also applied to the $(v,w)$ plane to ensure ray-like propagation in all three dimensions.

We now apply this methodology to design a metamaterial-based optical element, which can mold the light at sub-wavelength scale and in all three dimensions. This unusual manipulation of light is carried out by cascading multiple metamaterials-based elements with the ability to direct light or bend it by 90 degrees. The 90 degree *bending* elements are designed using a circular cylindrical coordinate transformation with rotation about the $w$-axis[23]

$$x + iy = \exp(u + iv) \qquad u \in (u_1, u_2), \, v \in (0, \pi/2) \qquad (4)$$

where $u_1$ and $u_2$ correspond to the internal and external radii of the bends (see also Fig. 1 in Supplementary Information). A uniaxial quasi-flat local dispersion enables the confinement of the sub-wavelength beam profiles both along $u$ and $w$ dimensions during propagation. Such EM material response can be obtained, e.g., by effective medium[15, 20, 21] composed of thin layers of alternating different materials (see blue and white layers in the inset of Fig2) such that

$s'_u = s'_w = (ps_1 + (1-p)s_2)$ and $s'_v = s_1 s_2 /((1-p)s_1 + ps_2)$. $s_1 = (\varepsilon_1, \mu_1)$ and $s_2 = (\varepsilon_2, \mu_2)$ represent the material response of the two media composing the layered structure, where $p$ is the filling fraction of layer 1.

Subsequently, we performed a full wave simulation, to demonstrate the applicability of our methodology in efficient design of small size optical elements (at the order of one vacuum wavelength). We have used COMSOL Multiphysics finite element solver to simulate three deep sub-wavelength scale beams flowing along three-dimensional paths in a metamaterials with effective EM response of $s'_u = s'_w = 0.005$ and $s'_v = -200$. Such effective EM response can be realized, e.g., using constitutional materials with $\varepsilon_1 = \mu_1 = 1$ and $\varepsilon_2 = \mu_2 = -0.99$ (see e.g., refs 14, 24) with equal filling fractions of 1/2. Fig 2 shows that the beams, at a width of $\lambda/10$, are clearly maintained throughout the three-dimensional propagation in the metamaterials. The beam size may be further confined by using metametarials with even stronger EM response by means, e.g., of EIT[25, 26], or metal-dielectric nano-spheres[27, 28].

The design methodology can also be exploited to develop an actual optical "transformer" that can project images of deep sub-wavelength objects to near or far field with negligible distortion. Here we show such a design of metamaterials-based optical element constructed by layers of metal and dielectric materials of realistic properties, forming two domains (I and II) of cylindrically symmetric curvilinear orthogonal coordinates (Fig. 3a). The curvilinear coordinates are mapped onto the Cartesian space ($x$, $y$, $z$) through bipolar transformations[23]

$$x - iy = R \coth(\frac{u_1 + iv_1}{2}) \quad \text{for region I } u_1 \in (-\infty, \infty), v_1 \in (\pi/2, \pi) \quad (5a)$$

and

$$y + ix = L - \sqrt{L^2 - R^2} \coth(\frac{v_2 + iu_2}{2})$$

$$\text{for region II } u_2 = 2\tan^{-1}\left(\tanh\left(\frac{v_{R,L}}{2}\right)\coth\left(\frac{u_1}{2}\right)\right), \quad v_2 \in (0, v_{R,L}) \quad (5b)$$

where $v_{R,L} = 2\tanh^{-1}(\sqrt{(L-R)/(L+R)})$, $L$ and $R$ are the length and inner radius of the element, respectively and the cylindrical symmetry imposing $z = w$ for both regions (See Supplementary Information Fig. 2). The ray trajectories conform to the curved $v$ axis (constant $u$) providing magnifying imaging with the relations

$$x_2 = mx_1, \quad m = 1 + \frac{L}{R} \quad (6)$$

where $x_1$ and $x_2$ are the lateral sizes of the object at the bottom plane, and image at the top plane of the element, respectively. The magnification factor $m$ is determined solely by the geometrical relation between $L$ and $R$, hence it is fully controlled by the design of the optical element.

We have simulated the actual imaging of a set of arbitrarily spaced sub-diffraction-limited objects at a size of $\sim \lambda/20$ in a composite of alternating Ag ($\varepsilon_{Ag} = -6.067 + 0.1967i$) and GaN ($\varepsilon_{GaN} = 6.088$) curved layers at operating vacuum wavelength $\lambda = 431$ nm (Fig. 3b). Here, the metal-dielectric interfaces comply with the *v=const* contours, and the local anisotropic response is determined by the effective medium properties $\varepsilon'_v = \varepsilon_{Ag}\varepsilon_{GaN}/(p\varepsilon_{GaN} + (1-p)\varepsilon_{Ag})$, $\varepsilon'_u = p\varepsilon_{Ag} + (1-p)\varepsilon_{GaN}$, and $\mu'_w = 1$ for TM polarization. By keeping the filling fraction as $p = 1/2$ over the entire space, the conditions of quasi-flat dispersion and slowly-varying effective material response are satisfied, enabling the use of ray tracing for imaging objects with sizes as small as $\lambda/20$.

The sub-diffraction limited object, shown in Fig. 4a, is imaged to the far-field under 5x magnification (blue line in Fig 4b). The numerical "control experiment" of light radiation in the absence of the metamaterials (i.e., free-space), results in a broad bell-shaped field distribution, with no resemblance to the original object (green line in Fig 4b). Namely, only a small fraction of the spatial information is transferred, whereas the fine features, carried by the evanescent waves are lost in free space. Fig. 4c depicts linear relations between the object and image positions, showing negligible spatial distortions.

In conclusion, we demonstrated a new approach of ray optics at the deep sub-wavelength scale, using metamaterials with properly designed dispersion characteristics. We developed a design methodology that combines quasi-flat dispersion and conformal mapping to obtain an equivalent space, represented by curvilinear orthogonal coordinates in which the material response is slowly varying. Using this methodology, we have shown a ray-like flow of light at scale of λ/10 in a metamaterial-based element. Sub-wavelength ray optics offers new exciting possibilities in science and technology applications such as high density integrated photonics, nanoscale microscopy, nanoscale optical tweezers and optical lithography.

## Methods

For numerical calculation of the three-dimensional molding of subwavelength light (Fig. 2), the metamaterial conditions for local quasi-flat dispersions were directly implemented in the simulation domain of the commercial finite element solver COMSOL Multiphysics. Regarding 90 degree bending elements, the rotational characteristics of the cylindrical curvilinear orthogonal coordinates ($u$, $v$, $w$) with regard to the global Cartesian coordinates ($x$, $y$, $z$) ($w$ axis is set to equal $z$ axis) enables representing the material responses by the global Cartesian coordinates

$$\begin{pmatrix} s'_{xx} & s'_{xy} & 0 \\ s'_{yx} & s'_{yy} & 0 \\ 0 & 0 & s'_{zz} \end{pmatrix} = \begin{pmatrix} \dfrac{s'_u x^2 + s'_v y^2}{x^2 + y^2} & \dfrac{(s'_u - s'_v)xy}{x^2 + y^2} & 0 \\ \dfrac{(s'_u - s'_v)xy}{x^2 + y^2} & \dfrac{s'_u y^2 + s'_v x^2}{x^2 + y^2} & 0 \\ 0 & 0 & s'_w \end{pmatrix}$$

In the case of the imaging device (Figs. 3 and 4), curved alternating layers of GaN and Ag were explicitly defined in the simulation domain with actual material parameters. In order to realize the local quasi-flat dispersions, local filling fraction $p$ was controlled to be near 1/2 by keeping the constitutional layers thickness constant (15 nm) along the $x = 0$ direction in Fig. 3b. In all simulations, scattering boundary condition has been applied at the external boundaries to absorb outgoing waves from the simulation domain.

**Figure legends**

**Figure 1: Extending ray optics to the sub-wavelength scale. a**, Molding light at sub-wavelength scale along three-dimensional arbitrarily curved ray trajectories in a metamaterial. **b**. curvilinear representation of the (*u,v*) axes. The *v*-axis (*u=const*) and *u*-axis (*v=const*) *contours* are represented by the dashed and solid grey lines, respectively. The light rays correspond to the *v*-axis trajectories. sub-wavelength diameter beams (represented by the yellow tubes) can be molded along these trajectories. **c**. Quasi-flat dispersion curves in Metamaterials, at an arbitrary point A in the Cartesian representation of the (*u,v*) mesh (green curve), and at the point B(A) (see Fig. 1b) in the curvilinear representation of (*u,v*) with respect to the global coordinates (*x,y*) (blue curve). These curves are compared to the dispersion curve of the vacuum, represented by the orange circle with a radius $k_0$.

**Figure 2 Three-dimensional bending of light flow at sub-wavelength scale.** Finite element calculation of three sub-wavelength scale (at width $\sim \lambda/10$) light beams flowing along three-dimensionally curved trajectories. The magnitude of the electric field |E| is presented by crosscuts at different orientations. The inset illustrates one way to obtain such light flow by a multi-layer design of the metamaterials; the blue and white layers correspond to constitutional alternating layers of materials with EM response of $(\varepsilon_1, \mu_1)$ and $(\varepsilon_2, \mu_2)$, respectively.

**Figure 3 New metamaterial-based optical transformer. a.** Scheme of the "optical transformer" consisting of two domains of alternating curved layers of Ag (pink) and GaN

(white). The *v*-axis in the field map is presented by the blue lines, while the *u*-axis complies with the interfaces between the layers. The light flows along the *v*-axis. **b**. Magnetic field intensity profile $|H|^2$ of arbitrarily spaced sub-diffraction-limited objects exemplifying the sub-wavelength imaging and magnification performed by the element.

**Figure 4 Magnification characteristics of the imaging devices. a**. Intensity distribution of the same arbitrarily spaced sub-diffraction-limited objects as in Fig 3 at the object plane. **b.** The same objects magnified by 5× at the image plane by the transformer ($R = 525$ nm, $L = 2115$ nm, blue line) compared with the propagation of the same objects in air (green line). **c**, Magnification characteristics (the ratio between the displacement of a point object at the image plane, and its displacement at the object plane) for two different transformers; 5× ($R = 525$ nm, $L = 2115$ nm, presented by the blue line) and 7× ($R = 405$ nm, $L = 2445$ nm, presented by the red line). The circular and square markers correspond to the numerical calculation results. The solid lines are the analytical calculation derived from equation (6).

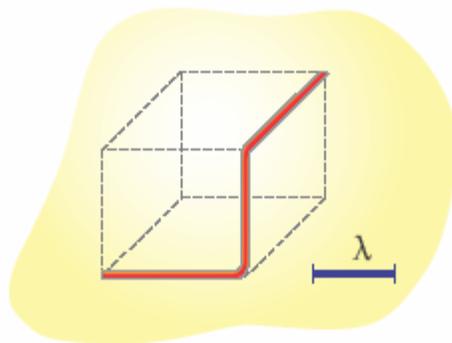

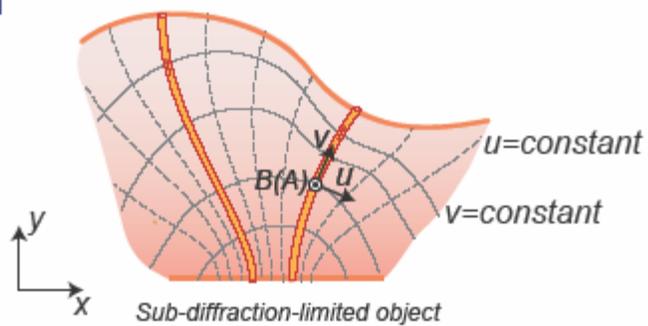

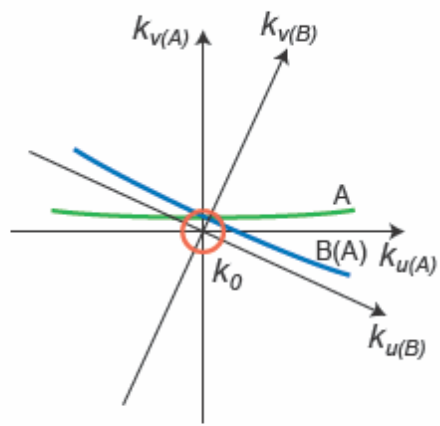

Fig. 1

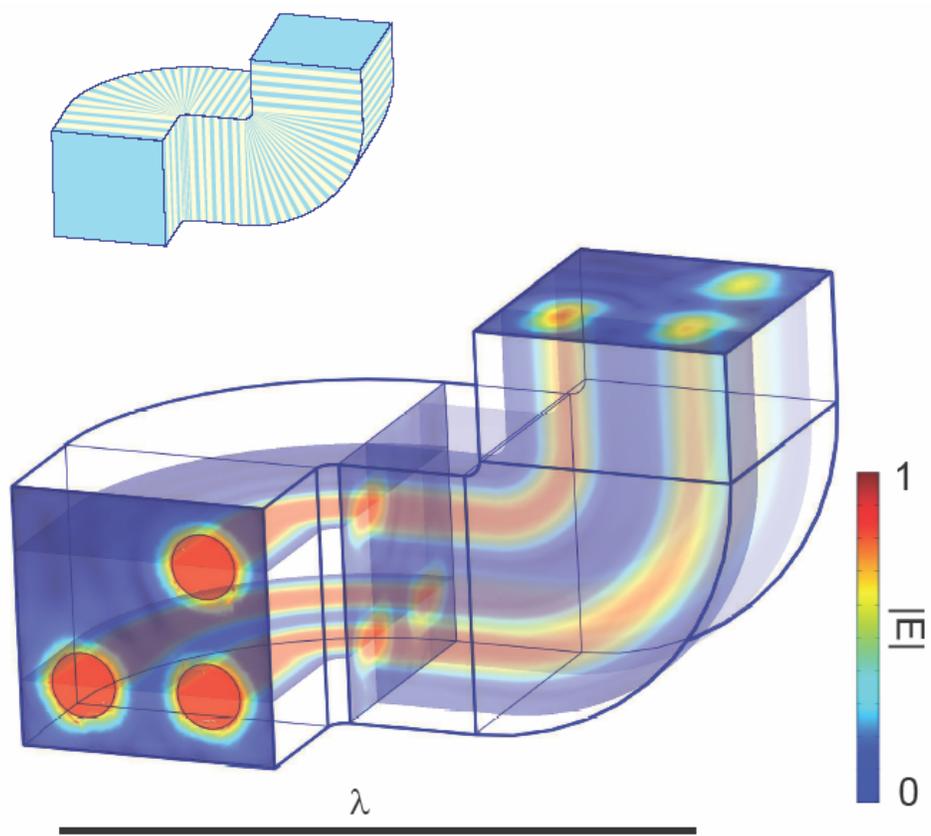

Fig. 2

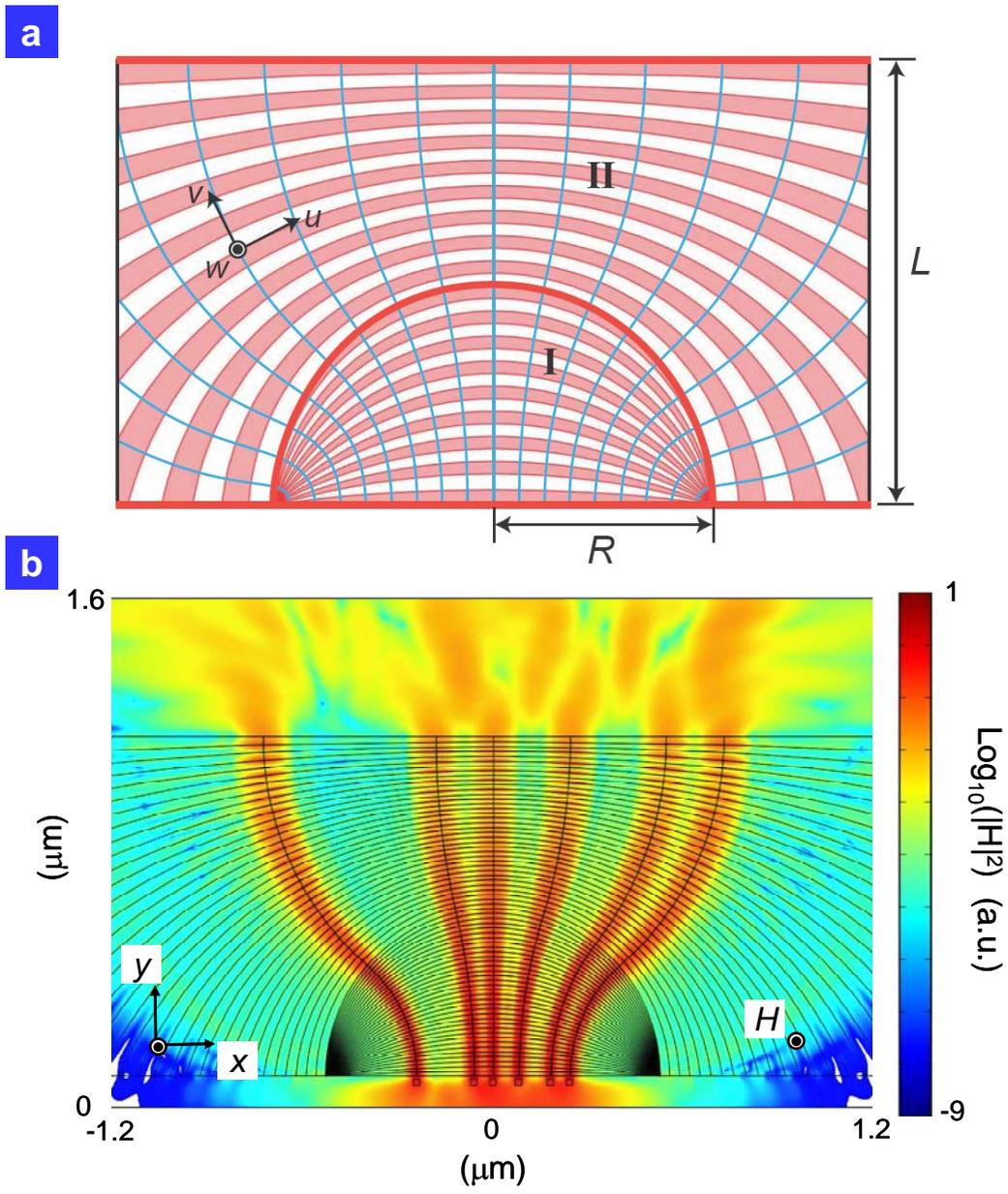

Fig. 3

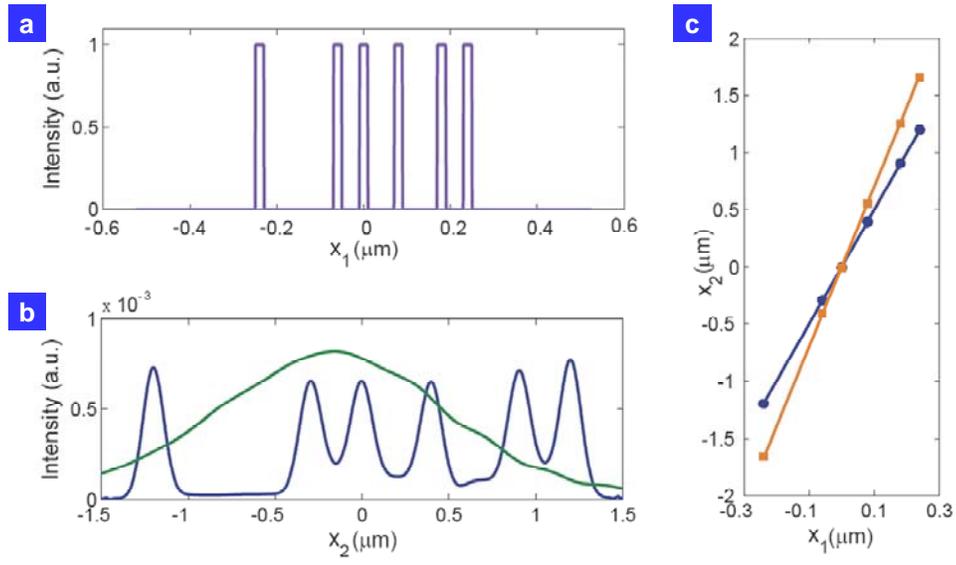

Fig. 4

# Supplementary Information

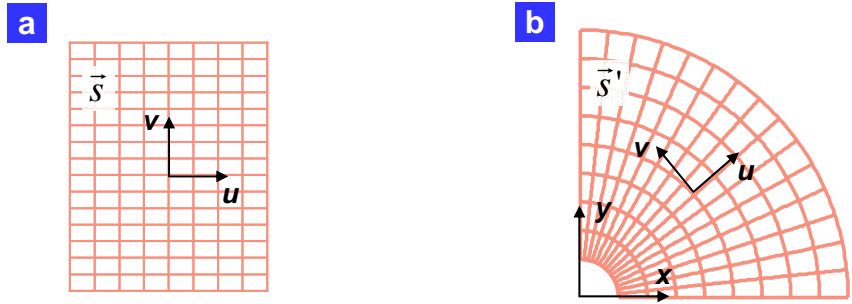

**Supplementary Figure 1 Schematic of conformal transformation of 90 degree bending optical element. a,** A Cartesian mesh ($u$, $v$) and **b,** its exponential function transformation

$$x + iy = \exp(u + iv)$$

enable a 90 degree bending optical element in the global ($x,y$) coordinates. With regard to the metamaterial properties in the coordinate transformation, the anisotropic response $\vec{s} = (\varepsilon, \mu)$ and $\vec{s}' = (\varepsilon', \mu')$ in each representation are related with each other as in equations (1) and (2). In Fig. 2, the quasi-flat dispersion condition is applied such that the diagonalized components of local response are kept constant (i.e., $s'_v = -200$ and $s'_u = s'_w = 0.005$) throughout the bending elements as shown in Supplementary Fig. 1b. In this case, outer ray trajectories experience longer optical path length (i.e., by propagation constant $k_v$ along $v$ direction in equation (3)) as proportional to $\sqrt{x^2 + y^2} \times \left( \sqrt{\varepsilon'_u \mu'_w} \text{ or } \sqrt{\varepsilon'_w \mu'_u} \right)$. This can be analyzed by transforming back to the simpler Cartesian mesh as in Supplementary Fig. 1a, where spatially dependent scaling factor $h_u^2 = h_v^2 = x^2 + y^2$ induces renormalization of effective phase index along propagation direction as $\sqrt{\varepsilon_u \mu_w} = \sqrt{\varepsilon_w \mu_u} = \sqrt{\varepsilon'_u \mu'_w} \sqrt{h_u h_v}$. It is a slowly varying index change satisfying the ray description of light flow.

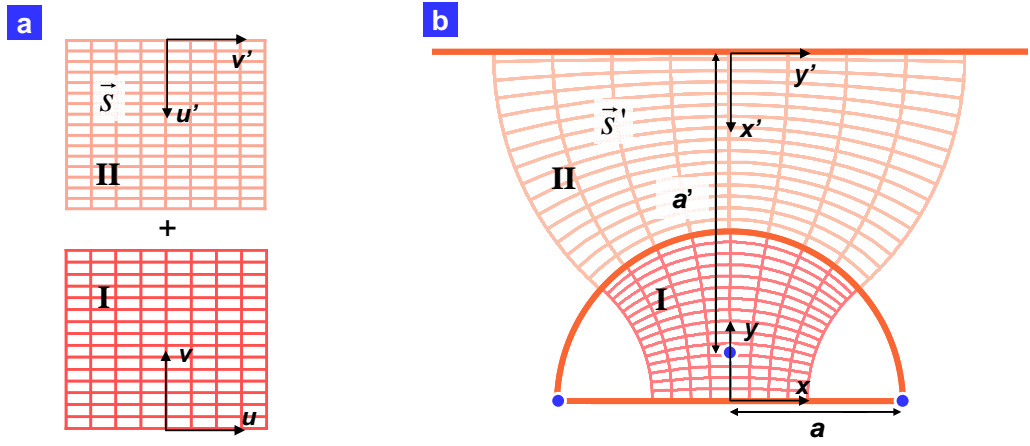

**Supplementary Figure 2 Schematic of conformal transformation of optical transformer. a,** Two Cartesian meshes ($u$, $v$) (region I) and ($u'$, $v'$) (region II) and **b,** their curvilinear orthogonal transformation by bipolar function in the ($x$, $y$) and ($x'$, $y'$) coordinates, respectively, constitute optical transformer. Bipolar function provides a complex plane transformation described by

$$x - iy = a \coth(\frac{u + iv}{2})$$

Note mesh orientation of the upper region II is rotated in comparison to the lower region I. In the manuscript, one global coordinates ($x$, $y$, $z$) is used (equation (5)) to describe the whole imaging device to make $v(x, y) = const$ curves conform to the interfaces between different constitutional material layers in Fig. 3a. The two regions I and II are designed to share a common half circle (orange line in the middle of Supplementary Fig. 2b) as their interface. The resulting curvilinear orthogonal coordinates (supplementary Fig. 2b) gives a field-map that is similar to an electrostatic problem, where different electrical potentials are enforced at the three electrodes (i.e., orange lines at the top, middle and bottom).